\documentclass[10pt]{article}

\usepackage{amsmath}
\usepackage{amssymb}

\usepackage{graphicx}

\usepackage{cite}

\usepackage[labelfont=bf,labelsep=period,justification=raggedright]{caption}

\makeatletter
\renewcommand{\@biblabel}[1]{\quad#1.}
\makeatother

\date{}

\newcommand{\be}{\begin{eqnarray}}
\newcommand{\ee}{\end{eqnarray}}

\newcommand{\ipred}{I_{\rm pred}}

\begin{document}

\begin{flushleft}
{\Large
\textbf{Integrated information increases with fitness in the evolution of animats}
}
\\
Jeffrey A. Edlund$^{1}$, 
Nicolas Chaumont$^{2,6}$, 
Arend Hintze$^{2,3,6}$,
Christof Koch$^{1,7,8}$,
Giulio Tononi$^4$,
Christoph Adami$^{2,5,6,\ast}$

\bf{1} Computation \& Neural Systems, California Institute of Technology, Pasadena, CA
\\
\bf{2}  Keck Graduate Institute of Applied Life Sciences, Claremont, CA 
\\
\bf{3} Computer Science and Engineering, Michigan State University, East Lansing, MI
\\
\bf{4} Department of Psychiatry, University of Wisconsin, Madison, WI
\\
\bf{5} Microbiology \& Molecular Genetics, Michigan State University, East Lansing, MI 
\\
\bf{6} BEACON Center for the Study of Evolution in Action,  Michigan State University, East Lansing, MI 
\\
\bf{7} Department of Brain and Cognitive Engineering, Korea University, Seoul Korea
\\
\bf{8} Allen Institute for Brain Sciences, Seattle, WA
\\
$\ast$ E-mail: adami@kgi.edu
\end{flushleft}

\section*{Abstract}
One of the hallmarks of biological organisms is their ability to integrate disparate information sources to optimize their behavior in complex environments. How this capability can be quantified and related to the functional complexity of an organism remains a challenging problem, in particular since organismal functional complexity is not well-defined. We present here several candidate measures that quantify information and integration, and study their dependence on fitness as an artificial agent (``animat") evolves over thousands of generations to solve a navigation task in a simple, simulated environment.  We compare the ability of these measures to predict high fitness with more conventional information-theoretic processing measures. As the animat adapts by increasing its ``fit" to the world, information integration and processing increase commensurately along the evolutionary line of descent. We suggest that the correlation of fitness with information integration and with processing measures implies that high fitness requires both information processing as well as integration, but that information integration may be a better measure when the task requires memory. A correlation of measures of information integration (but also information processing) and fitness strongly suggests that these measures reflect the functional complexity of the animat, and that such measures can be used to quantify functional complexity even in the absence of fitness data.

\section*{Author Summary}
Intelligent behavior encompasses appropriate navigation in complex environments that is achieved through the integration of sensorial information and memory of past events to create purposeful movement. This behavior is often described as ``complex", but universal ways to quantify such a notion do not exist.  Promising candidates for measures of functional complexity are based on information theory, but fail to take into account the important role that memory plays in complex navigation. Here, we study a different information-theoretic measure called ``integrated information", and investigate its ability to reflect the complexity of navigation that uses both sensory data and memory. We suggest that measures based on the integrated-information concept correlate better with fitness than other standard measures when memory evolves as a key element in navigation strategy, but perform as well as more standard information processing measures if the robots navigate using a purely reactive sensor-motor loop. We conclude that the integration of information that emanates from the sensorial data stream with some (short-term) memory of past events is crucial to complex and intelligent behavior and speculate that integrated information--to the extent that it can be measured and computed--might best reflect the complexity of animal behavior, including that of humans.

\section*{Introduction}
Complexity is visible in most scientific disciplines: mathematicians, physicists, biologists, chemists, engineers and social scientists all developed measures to characterize the complexity that they perceive in their systems, borrowing tools from each other but rarely if ever agreeing on a measure that could be used by all of them. Because the objects that each of these disciplines are most concerned with are so different, ranging from mathematical problems and computer programs over physical, chemical, or biological structures, to systems and networks of interacting agents, a convergence of quantitative measures of complexity is perhaps not likely. However, a universal framework that would be capable of adapting its notion to the specific discipline it is applied to would be a welcome trend.
Complexity measures abound, but exhaustive reviews are few. A good introduction to the dynamical systems approach to complexity is Ref.~\cite{badii-politi-97}, but it does not cover biological applications. The overviews~\cite{adami-02,Adami2004,Adami2009} focus on the complexity of biological sequences but not on their structure, and mostly ignore the complexity of networks. Neural complexity measures are reviewed in~\cite{Sporns2011}. 

Among the different measures of complexity, some attempt to quantify the structure~\cite{Lofgren1977,Chaitin1979,Papentin1980,Papentin1982,ThomasReif1993,Thomasetal2000,SoloveichikWinfree2006,Ahnertetal2010}, others the sequence giving rise to that structure~\cite{Kolmogorov1965,LempelZiv1976,EbelingJimenezMontano1980, LiVitanyi1997,GellMannLloyd1996,AdamiCerf2000}, and others again the function of the sequence or system~\cite{McShea2000,Szostak2003,Hazenetal2007}.  All these studies attempt to capture ``that which increases when self-organizing systems organize themselves"~\cite{Bennett1995} (a non-exhaustive list is presented in Ref.~\cite{Lloyd2001}). Increasingly, measures based on information theory are being used to quantify the complexity of living systems, because information provides its owner an obvious fitness advantage compared to those without information by conferring the ability to make predictions about the environment they operate in~\cite{Tayloretal2007,Polani2009,RivoireLeibler2011}.  In particular Rivoire and Leibler~\cite{RivoireLeibler2011} study statistical measures based in information theory that maximize the fitness of agents that respond to variable environments, but they do not study evolution. Information-theoretic measures of complexity are reviewed in~\cite{Bonchev2009} and applications to graphs in~\cite{DehmerMorshowitz2011}.

Here, we study how information-theoretic measures of complexity could be applied to capture the complexity of nervous systems~\cite{KochLaurent1999,Sporns2011}, or more generally speaking, any structure controlling a perception-action cycle. In the absence of any well accepted definition of complexity, we study the correlation of different measures to {\em organismal fitness}, following the intuition that a well-defined measure of control structure complexity should increase during adaptation~\cite{McShea2000}. Fitness is a quantitative measure that predicts the long-term success of a lineage~\cite{Haldane1932,MaynardSmith1969}, and is given by the {\em expected number of offspring} of an average representative with the given genotype. Unfortunately, this is only a quantitative measure for the simplest of organisms where the expected number of offspring can be determined from the replication rate, or in direct competition experiments (see, e.g.,~\cite{Lenskietal1991}). For more complex organisms, relative fitness can only be estimated in hindsight, and cannot be used as a proxy for organism complexity. However, if we evolve control structures {\it in silico} where complete fitness information is available, we can use fitness (within a niche) as an independent arbiter of putative information-based measures of complexity: any measure that does not increase as the organism learns to exploit its environment is unlikely to reflect complex information processing. Because in this type of evolution experiment the number of offspring is directly proportional--on average--to the {\em performance} of the organism in a task critical to its survival, we here study the correlation of complexity directly with performance or function.

Note that because fitness necessarily refers to the environment (it measures how well the organism ``fits" its niche by exploiting the niche's attributes), fitness cannot be used to compare organism complexity across niches (such as attempting to compare an elephant and an ant in terms of their fitness), but it does reveal functional differences between types that are due to efficiencies of exploiting the same environment. For biological organisms that occupy the same niche, that is,  ``make a living" in the same manner, relative fitness should correlate with relative functional complexity.  Is it true that given a constant environment the more complex organism is necessarily more fit? Answering this question in the affirmative clearly biases our notion of complexity: only useful characters are deemed complex, useless ones are not. While such a bias may be restrictive for structural complexity, it is not so for information-theoretic measures of complexity, as information (if it can be used to reduce uncertainty) will {\em always} be useful: if it were not, it should be called entropy instead~\cite{Tayloretal2007,Donaldson-Matascietal2010}.

\subsection*{Predictive information}
Perhaps the best known information-based measure of functional complexity is ``predictive information"~\cite{Bialeketal2001}, which quantifies the amount of information that can be extracted from sensorial data in order to select actions that are useful to the organism. In this manner, predictive information is able to separate out those features of the sensorial data that are relevant for behavior, and quantifies the amount of information processed by the organism. Predictive information has also been proposed as a measure of complexity of function~\cite{Bialeketal2001}.

If we describe a control network's input variables (``sensors'', or ``stimuli") at time $t$ by the random variable $S_t$ and the output variables (``motors'', or ``response") at that time by $R_t$, then the shared information (used for prediction)
is~\cite{Bialeketal2001} 
\be
I_{\rm pred}&=&I(S_t:R_{t+1})=H(R_{t+1})-H(R_{t+1}|S_t)\nonumber \\
&=&\sum_{s,r}p(s_t,r_{t+1})\log \frac{
  p(s_t,r_{t+1})}{ p(s_t) p( r_{t+1} )} \label{ipred}\;,
\ee
where ${\rm Pr}(S_t=s_t)\equiv p(s_t)$ and ${\rm Pr}(R_t=r_t)\equiv p(r_t)$ are the probability distributions of the sensor and response variables at time $t$, respectively, and $p(s_t,r_{t+1})$ is the joint probability distribution of the sensor and response variables ``in the future and the present"~\cite{Bialeketal2001} (we use the binary logarithm throughout and assume that the network evolves along discrete time steps). $\ipred$ characterizes the capacity of the control system to predict the future one time step ahead, using the present sensorial information. Essentially, it quantifies the correlation between inputs and outputs,  and can be thought of as the Kullback-Leibler divergence (or relative entropy) between the full probability distribution $p(s_t,r_{t+1})$ and the product of the independent ones, $p(s_t)p(r_{t+1})$.

Note that for Markov processes, the one-step shared entropy (\ref{ipred}) is equal to the shared entropy between the entire past and the entire future (see~\cite{Ayetal2008}, Appendix A.1), while this is not true for processes that can use memory. Predictive information was previously used to characterize the complexity of autonomous robot behavior without memory in Ref.~\cite{Ayetal2008} (see also Text S1). If the control structure is not purely reactive and uses information encoded in internal nodes to integrate sensorial information streams, we will need complexity measures that move beyond predictive information~\cite{RivoireLeibler2011}.

\subsection*{Integrated information}

A fundamental and unique design principle of nervous systems is their
extraordinary degree of integration among highly-specialized
modules~\cite{FellemanvanEssen1991,Hagmannetal2008,Sporns2011}. Functional integration is achieved by an extended network of
intra- and inter-areal connections, and is reflected in dynamically shifting patterns of synchronization.  A precise way to measure a
system's capacity to integrate information was developed recently~\cite{tononi-08,balduzzi-tononi-08}, and applied to small, simple example networks. This measure, called $\Phi$ and measured in bits, is based on the notion that information integration is achieved by architectural designs that give rise to a single, functionally unified complex (high integration) while ensuring that such a complex has a very large repertoire of discriminable states (high information). $\Phi$ captures to what extent, informationally, the whole is more than the sum of its parts,
and cannot therefore be reduced to those parts. In this sense, $\Phi$ represents the synergy of the system.  Before introducing $\Phi$ proper, we define a few related quantities.

In order to study information integration, we have to define the 
information processed by the entire network, not just the sensors and motors as in Eq.~(\ref{ipred}). Let us represent the system as a joint random variable $X=X^{(1)}X^{(2)}\cdots X^{(n)}$, where the $X^{(i)}$ represent the {\em elements} of the system (the nodes of a control structure, such as a neuronal network). The random variable $X$ evolves as the system progresses forward in time, i.e., $X(t=0)\equiv X_0\to X_1\to X_2\to\cdots X_t$, and each variable $X_t$ is described by a probability distribution $p(x_t)$ to be found in states $x_t$ (here, we will restrict ourselves to binary random variables). At the same time, each node $i$ of the system has a time progression $X_0^{(i)}\to X_t^{(i)}$, and each variable $X_t^{(i)}$ is described by a probability distribution $p(x_t^{(i)})$. In the following, we formally define measures of information integration through $t$ time steps (from $0\to t$), but later focus on the computationally more accessible integration through a single average step from $t\to t+1$. 
 
The amount of information that is processed by the {\em entire system} through $t$ time steps is given by
\be
I(X_0: X_t)= \sum_{x_0,x_t}p(x_0,x_t)\log\frac{p(x_0,x_t)}{p(x_0)p(x_t)}\;. \label{info} 
\ee
where $p(x_0)$ and $p(x_t)$ are the probability distributions of the system at time $t=0$ and $t$ respectively, and $p(x_0,x_t)$ is their joint distribution. This measure reduces to the predictive information Eq.~(\ref{ipred}) for Markov processes connecting only sensor and response nodes, that is, if there are no internal (or hidden) variables. 

One way to measure information integration is to ask how much information is processed by the system above and beyond what is processed by the individual nodes or groups of nodes (modules). To do this, we introduce a {\em partition} of the network into $k$ parts, $P=\{P^{(1)},P^{(2)},\cdots,P^{(k)}\}$, where each $P^{(i)}$ is a part of the network: a non-empty set of nodes with no overlap between parts that completely tile the network. We can then define a quantity that measures how much the information processed by the entire network is more than the information processed by all the parts in this particular partition as follows.

Let $I(P_0^{(i)}:P_t^{(i)})$ be the information processed by the $i^{\rm th}$ part as the system progresses from time 0 to time $t$. Then, the synergistic information $SI$ processed by the network $X$ given a partition $P$ quantifies the extent to which the entire processed information is a sum of the information processed by the system's parts, and is
calculated as:
\be
SI(X_0\to X_t|P)=I(X_0: X_t)-\sum_{i=1}^k I(P_0^{(i)}:P_t^{(i)})\;. \label{SI}
\ee 
From an information-theoretic point of view, the synergistic information measures the excess amount of information that can be encoded in a ``multiple access" channel with correlated sources and a joint decoder~\cite{SlepianWolf1973} over and above what each of the individual channels (the parts of the partition $P^{(i)}$) can encode separately.
A measure related to the synergistic information is the ``effective information" $EI$:
\be
EI(X_0\to X_t|P)=\sum_{i=1}^kH(P_0^{(i)}|P_t^{(i)})-H(X_0|X_t)\;. \label{EI}
\ee
Here, $H(P_0^{(i)}|P_t^{(i)})$ is the conditional entropy of partition $P_0^{(i)}$ given the state of that partition $t$ time steps later, and $H(X_0|X_t)$ is the conditional entropy of the entire system $X$ at time step $t=0$ given the state of that system $t$ steps later (see also Text S3).  
The quantity (\ref{EI}) is the average over network states at time $t$ (states $x_t$) of the quantity called the ``effective information across a partition $P$" in Ref.~\cite{balduzzi-tononi-08}. If the probability distribution governing $X_0$ is uniform (maximum entropy), the two measures agree: $SI(X_0\to X_t|P)=EI(X_0\to X_t|P)$, but they  are different in general (see Text S3). Below, we will mostly use Eq.~(\ref{EI}).

In order to determine how a network integrates information, we should look for a partition that {\em minimizes}~(\ref{EI}), because it is easy to find a high value of $EI$ by assigning different parts to nodes that are strongly correlated. In essence, looking for the partition that minimizes $EI$ is tantamount to searching for the groups of nodes that are separated from other groups of nodes by a weak informational link~\cite{balduzzi-tononi-08}. To find this partition, expression (\ref{EI}) needs to be normalized because otherwise the partition that minimizes (\ref{EI}) will almost always be the one that divides a network of $N$ parts into one with $N-1$ parts and a single other node~\cite{balduzzi-tononi-08}. We define the ``Minimum Information Partition'' (or `MIP') as that partition that minimizes a {\em normalized}  $EI$:
\be
{\rm MIP_0}&=&\arg \min_P \frac{EI(X_0\to X_t|P)}{(k-1) \min_i\left[H_{\rm max}(P_0^{(i)})\right]}\;,
\ee
where $H_{\rm max}(P_0^{(i)})$ is the maximum entropy of the $i$th partition $P_0^{(i)}$. If the neurons are binary, then $H_{\rm max}(P_0^{(i)})$ is just the number of neurons in partition $i$. Armed with this definition of the MIP, our measure of information integration is:
\be
\Phi_0&=&EI(X_0 \to X_t|P={\rm MIP_0})\;.\label{t10} 
\ee
Note that $\Phi_0$ represents the average (over all possible final states of the network) of the state-dependent quantity $\Phi(x_t)$ defined previously~\cite{balduzzi-tononi-08}, and the subscript $0$ reminds us that the integration is measured from an initial probability distribution at time $t=0$ that is uniform.  

The measure can be adapted to characterize the information integration across a single time step simply by defining
\be
\Phi&=&EI(X_t \to X_{t+1}|P={\rm MIP}) \;\label{t1} 
\ee
with a commensurately defined MIP:
\be
{\rm MIP}&=&\arg \min_P \frac{EI(X_t\to X_{t+1}|P)}{(k-1) \min_i\left[H_{\rm max}(P_t^{(i)})\right]}\;,
\ee
where $H_{\rm max}(P_t^{(i)})$ is the maximum entropy of the $i$th partition at time step $t$. Note that we have omitted an index $t$ to $\Phi$ as defined in Eq.~(\ref{t1}) as we assume that for large $t$ $\Phi$ becomes stationary: $\Phi_t\to\Phi\;\;(t\to\infty)$.
This MIP, just as the one defined by Balduzzi and Tononi~\cite{balduzzi-tononi-08}, divides the network into disjoint parts that are maximally informationally disparate--those parts that are most independent.  As defined here, $\Phi$ is equivalent to the recently defined $\tilde\Phi_E$\cite{BarrettSeth2011}, because $EI(X_t\to X_{t+1}|P={\rm MIP})$ is based on the reduction (at time step $t+1$) in the Shannon entropy based on the empirical entropy at time step $t$, not on the reduction from the maximum entropy at time step 0 as in~\cite{balduzzi-tononi-08}. Thus, our Eq.~(\ref{t1}) is equivalent to Eq.~(29) in Ref.~\cite{BarrettSeth2011} (with $\varphi$ replaced with $\tilde\varphi$ ), except that we search all partitions rather than just bi-partitions, and the normalization factor of Barrett and Seth uses the largest of the actual entropies of the parts.   Because we will measure information integration for time series generated by a moving animat, we will use Eq.~(\ref{t1}) to quantify the animat's complexity in what follows.

If networks are small, it is possible to find the MIP by brute-force testing all possible partitions.  The number of partitions of $n$ nodes is the $n$th Bell number, $B_n$~\cite{rota-64}. Searching across all partitions is exceedingly expensive and scales faster than exponential.  For example, $B_3$=5, $B_{10}$=115,975, and $B_{16}\approx1.05\times 10^{10}$. For networks of realistic size, search heuristics will be the only way to find the MIP: for the nematode {\it C. elegans}, for example, $n=302$~\cite{KochLaurent1999}, and the number of partitions of this
network is the absurdly large number $B_{302}\approx 4.8\times 10^{457}$. Here, the largest networks we analyze have 12 nodes, but we have been able to calculate $\Phi$ for networks with up to 18 nodes using a fast exact algorithm that does not store all the partitions.

\subsection*{Main complex} \label{section:mainComplex} 
A system composed of a large network together with a single disconnected unit will always have $\Phi=0$, because minimizing
over all partitions finds the informational disconnect between the network and the disconnected node, and the minimum effective information between these parts is zero~\cite{balduzzi-tononi-08}.  A measure that captures information processing that is synergistic without being trivial can be obtained by defining the network's computational  {\em proper complex}~\cite{Tononi2010} as a subset $S$ of (joint) random variables within the system $X$ ($S \in X$) that maximizes $\Phi$ over all subsets {\em and} supersets, that is: \\
  
\begin{figure}[!ht]    
\begin{center}
\includegraphics[width=2.5in]{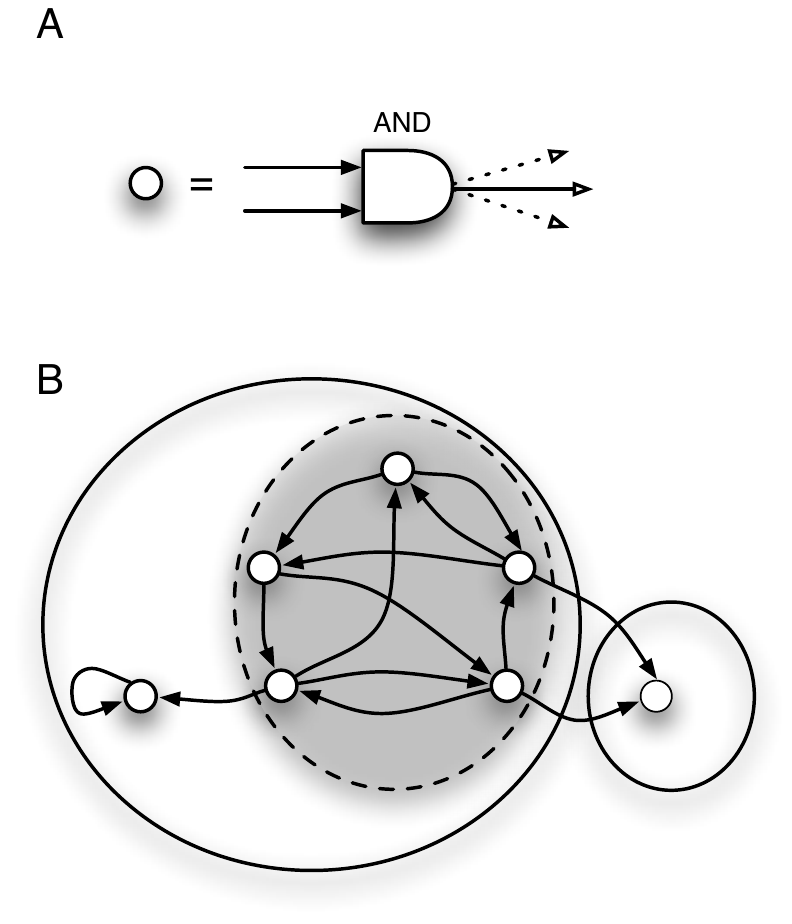}
\end{center}
\caption{{\bf Exemplar MIP and main complex.} {\bf A}: The logical units are AND gates with multiple outputs (each output
  is the AND of the two inputs). {\bf B}: A network of seven such units (877 distinct possible partitions).
  The  ${\rm MIP}$ for the entire system (solid lines) is a bi-partition, and the main complex (dashed line, shaded
  area) consists of five units. We compute $\Phi=0.269$ bits for the entire network, while $\Phi_{\rm MC}=1.327$ bits.}
\label{fig:didactic}
\end{figure}

 \noindent  If $\Phi(S)$ is defined as the $\Phi$ of subset $S$, then 
\begin{equation}
S \subseteq X\  \text{is a proper complex if}
\begin{cases}
	\Phi(S) \geq \Phi(R)  \ \forall \  R \subset S \\
	\Phi(S) > \Phi(T) \ \forall \ T\ \supset S\;.  
\end{cases}
\end{equation}
Each network can have several (proper) complexes, with smaller complexes of higher $\Phi$ embedded within larger complexes of lower $\Phi$. We define the {\em proper main complex} as the subset associated with largest $\Phi$ values over all subsets of the entire system.
We denote the information integration in the proper main complex as $\Phi_{\rm MC}$. A simple network with MIP and
main complex identified is shown in Fig.~\ref{fig:didactic}. 
\subsection*{Other integration measures}

Among all possible partitions, the ``atomic partition" that partitions the network into its individual nodes, plays an important role. For example, we can define the information processed by the network above and beyond the information processed by the individual nodes as
\be
SI_{\rm atom}=I(X_t:X_{t+1})-\sum_{i=1}^{n}I(X_t^{(i)}:X_{t+1}^{(i)})\;, \label{siatom}
\ee
where the first term is the total processed information $I_{\rm total}$, defined as
\be
I_{\rm total}=I(X_t:X_{t+1})=H(X_{t})-H(X_{t}|X_{t+1}) \;, \label{I_tot}
\ee
The negative Eq.~(\ref{siatom}) has previously been used to quantify the redundancy of information processing of a neural network~\cite{Atick1992,NadalParga1994}, see also~\cite{Schneidmanetal2003a}. Incidentally, Barlow has long argued that reducing redundancy (and thus compressing the sensorial information stream maximally) is the main purpose of the structure of the sensorial information-processing system~\cite{Barlow1961}, and we would then, if $I_{\rm total}$ is fixed, expect a maximization of fitness to go hand-in-hand with a minimization of $SI_{\rm atom}$ and therefore a maximization of redundancy,  

$I_{\rm total}$  measures the shared entropy between the system at adjacent time points, and is a useful measure to determine whether an increase in $\Phi$ is due solely to increased information processing by the entire network (resulting in an increased $I_{\rm total}$) rather than the effective integration of that information.
Writing $I(X_t^{(i)}:X_{t+1}^{(i)})=H(X_{t}^{(i)})-H(X_{t}^{(i)}|X_{t+1}^{(i)})$  for each node $i$, we see that
\be
SI_{\rm atom}= -H(X_{t}|X_{t+1})+\sum_{i=0}^nH(X_{t}^{(i)}|X_{t+1}^{(i)})-{\cal I} \;, \label{iatom}
\ee
where $n$ is the number of individual nodes in the network and where
\be
{\cal I}=\sum_{i=1}^nH(X_{t}^{(i)})-H(X_{t})\;.\label{integ}
\ee
This quantity has been called ``multi-information"~\cite{McGill1954b,Schneidmanetal2003a}), and was used as a measure of brain complexity
called ``integration" in~\cite{Tononietal1994,Lungarellaetal2005,LungarellaSporns2006}, where the sum was over the components of a network rather than the nodes. Thus, ${\cal I}$ is an ``atomic" form of the Tononi-Sporns-Edelman (TSE)-complexity~\cite{Tononietal1994}. 
Note that none of the measures discussed in this section should depend on $t$ if $t$ is large enough because we assume that at large times the probability distribution $p(X_t)$ becomes stationary.

The first part in Eq.~(\ref{iatom}) is nothing but the effective information $EI$ (\ref{EI}), but for the ``atomic partition", that is, the partition where each part is given by the individual nodes in the entire network and for $t\to t+1$. Thus, 
\be
 SI_{\rm atom}= \Phi_{\rm atom}-{\cal I}\;, \label{isum}
 \ee
where
 \be
 \Phi_{\rm atom} =EI(X_t \to X_{t+1}|P=P_{\rm atom})= \sum_{i=0}^nH(X_{t}^{(i)}|X_{t+1}^{(i)})-H(X_{t}|X_{t+1})\;. \label{atom}
\ee
Eq.~(\ref{atom}) may be a particularly useful measure to approximate $\Phi$ when a search for MIPs is computationally infeasible. It has previously been introduced under the name ``stochastic information" by Ay~\cite{Ay2001,AyWennekers2003a,AyWennekers2003b}.
However, it is neither an upper nor a lower bound on $\Phi$. Because of its construction ($\Phi_{\rm atom}=SI_{\rm atom}+{\cal I}$), it incorporates elements of information processing (the excess information processed, in time, by the system above and beyond the information processed by each of the nodes) as well as integration. In other words, $\Phi_{\rm atom}$ encompasses both temporal and spatial synergies of the network.

\section*{Results}

In order to test how different measures of functional complexity change as a system adapts to function in its world, we evolve
controllers for animats~\cite{Wilson1991} that have to solve a task that requires sensory-motor coordination as well as memory. 
Ay and coworkers tested  predictive information Eq.~(\ref{ipred}) as a measure of system complexity when evolving a simulated autonomous robot to solve a simple maze, and found that $\ipred$ reflects the performance of the robot~\cite{Ayetal2008}. Lungarella and coworkers used information-based complexity measures to understand how appropriate motor action of embodied agents shapes the signal structure perceived by the agent's sensors~\cite{Lungarellaetal2005}, and studied the information flow through the control structures~\cite{LungarellaSporns2006}. Klyubin and coworkers used mutual information between an agent's starting position and a representation of this information in the agent's memory to evolve sensorimotor control structures, and used measures of synergy to study whether the positional information could be factorized within the sensors~\cite{Klyubinetal2007}. 

\subsection*{Description of evolutionary system}
\begin{figure}[!ht] 
   \centering
   \includegraphics[width=2in]{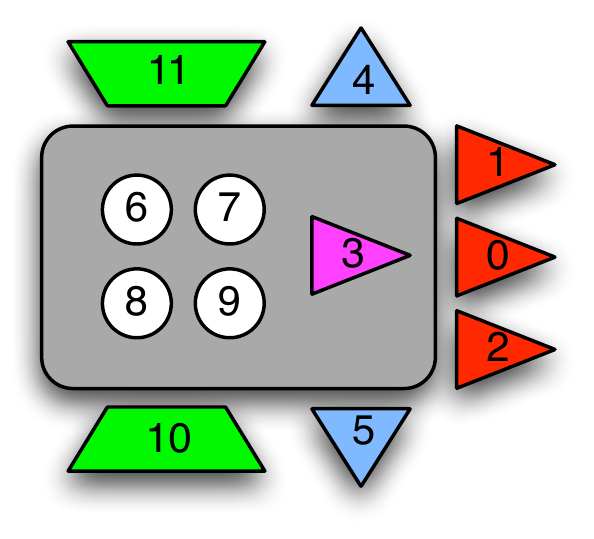} 
   \caption{{\bf Embodied virtual agent (animat) with six sensors, two actuators, and four internal nodes.} The complete animat is described by 12 bits: three front sensors (red triangles; \# 0,1 \& 2), two lateral collision detectors (blue triangles; \# 4 \& 5), and a single ``door" sensor  (magenta, \#3) that relays the direction of the next opening in the maze (but only while standing in the door).
     The actuators (trapezoids; \# 10 \& 11) encode the actions ``move left, move right, move forward, do nothing". The internal nodes (circles; \# 6-9) can potentially store states used for internal processing.}
   \label{fig:robot}
\end{figure}

Our animats are embodied controllers with six binary sensors and two (binary) actuators, as well as four internal bits that can be used for memory or processing (Fig.~\ref{fig:robot} and Methods). 

The controllers are stochastic Markov networks (see, e.g.,~\cite{KollerFriedman2009}), that is, networks of random variables with the Markov property, where edges between nodes encode arbitrary fuzzy logic gates. As such, the edges could represent simple binary logic gates or more complex computational units. Because these networks actually encode {\em decisions}, strictly speaking they are encoding discrete-time stochastic Markov decision processes (MDPs). Fundamentally, our Markov networks are related to the hierarchical temporal memory (HTM) model of neocortical function~\cite{HawkinsBlakeslee2004,GeorgeHawkins2005,GeorgeHawkins2009} and the HMAX algorithm~\cite{RiesenhuberPoggio1999}, except that the organization of our stochastic Markov networks need not be strictly hierarchical because it is determined via genetic evolution rather than top-down design (see Methods). 

\begin{figure}[!ht] 
   \centering
   \includegraphics[width=0.75\linewidth]{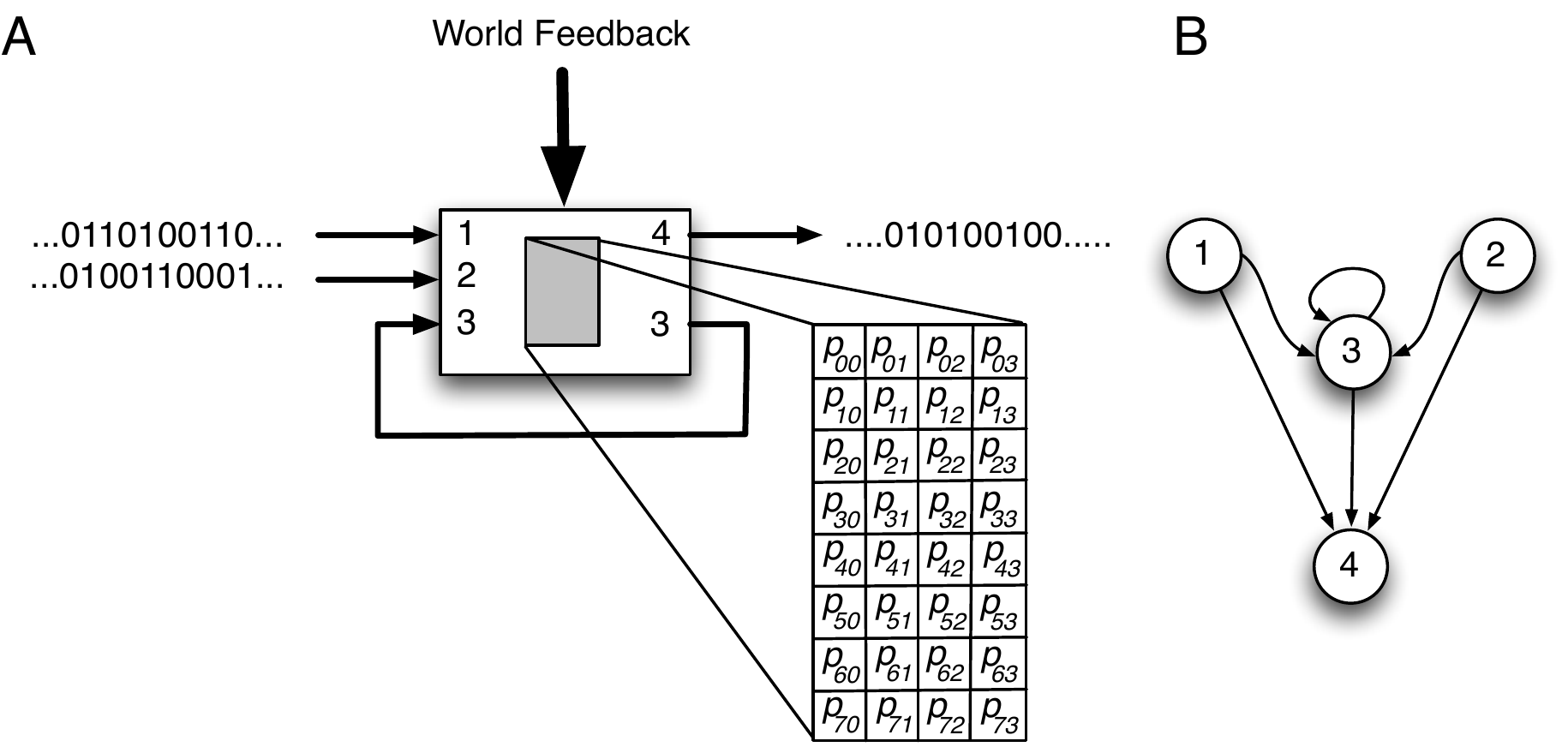} 
  \caption{{\bf Hidden Markov Gate representation.} {\bf A}: An HMG with three binary input and two output Markov variables, where one of the outputs is fed back into the HMG (a hidden variable). The state transition table has $2^3\times2^2$ entries that are determined by genetic evolution (see Methods and Text S2). In the gate shown, bit three is a hidden state and can be used to implement a one-bit memory. In principle, the probabilities in the HMG transition table can also be tuned via reinforcement learning using a signal from the environment (``World Feedback"). However, this capacity is not utilized in the present work.
 {\bf B}: The ``dual" representation of this gate, where the Markov variables are nodes, and the gate connects these via edges. This network is obtained by drawing a directed edge between bits that affect each other causally via the logic gate. Because bit 3 feeds back to itself, for example, it is given the same identifier and there is a directed arrow from bit 3 to itself as well as bit 3 to bit 4. See Text S2 and Fig. S1 for details on the genetic encoding and network visualization of HMGs.}
   \label{fig:hmg}
\end{figure}

In what follows, the edges connecting the random variables are implemented as  {\it Hidden Markov Gates} (HMGs). Each such gate is a
probabilistic finite state machine defined by its input/output structure and state transition probabilities (see Fig.~\ref{fig:hmg}A). For example, if `100' was applied to the input state of the HMG in Fig.~\ref{fig:hmg}A, `11' is the output with probability $p_{43}$ [$P(100\to11)=p_{43}$], while an input `111' generates `01' with probability $p_{71}$ [$P(111\to01)=p_{71}$], and so forth. Such a gate can also be represented as its dual graph, where the signal lines become the nodes of the Markov network, and the edges between them represent the computation performed by the HMG (Fig.~\ref{fig:hmg}B). In this representation, arrows indicate causal influence via an HMG, so in Fig.~\ref{fig:hmg}B for example, variable 4  is influenced by variables 1,2, and 3 (as is variable 3), while variables 1 and 2 only have outgoing arrows: they only influence variables 3 and 4 but are not affected by any other variable. 

\begin{figure}[h]
\begin{center}
\includegraphics[width=\linewidth]{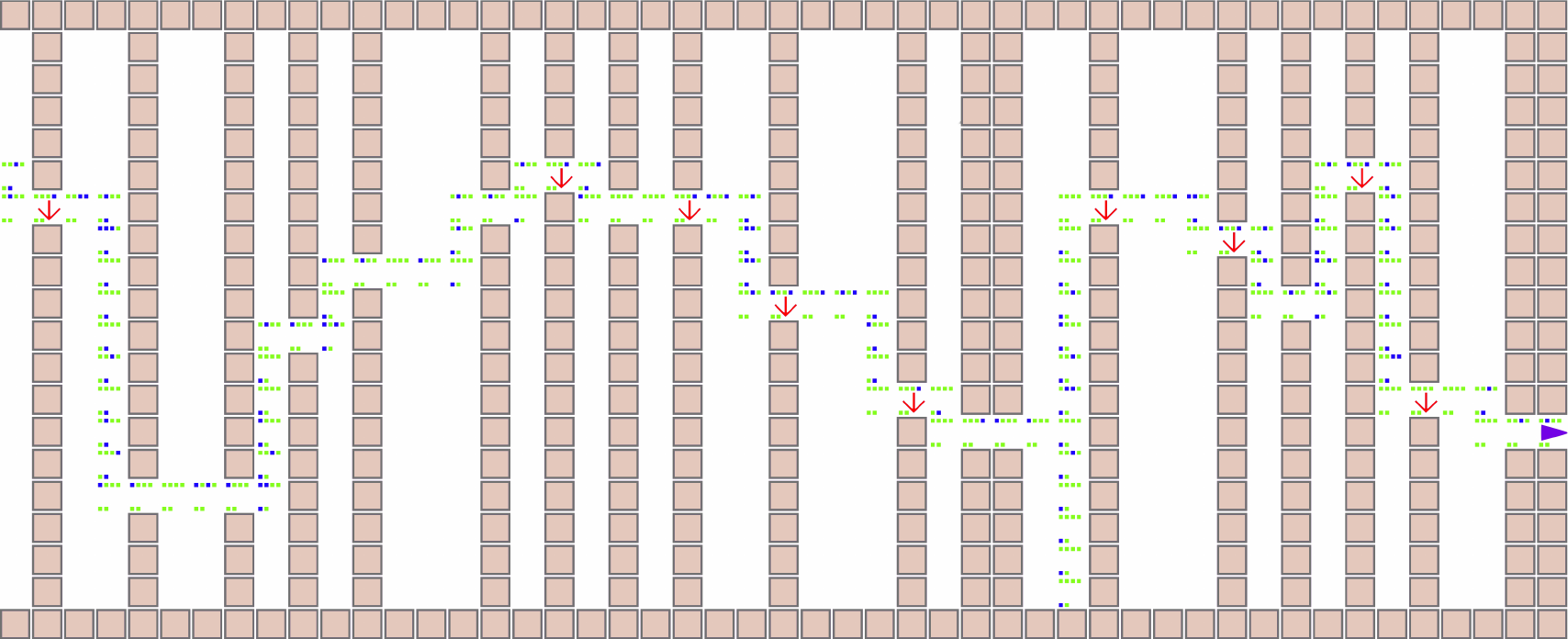} 
\end{center}
\caption{{\bf Maze structure and animat trajectory.} Part of one of the test mazes, along with the trajectory of an adapted animat as well as a view of the animat's brain (the four internal nodes 6-9, top four pixels in each animat location) and the motor outputs (bottom two pixels). A bit set to `1' is indicated in green, while blue  indicates a bit set to `0'. The value of the sensory bits can be inferred from the animat's location. The downward pointing arrow inside a door reminds us that the animat would perceive a `1' on its door sensor at that location (indicating that the next door will be found to the right of the animat's position). If the door is straight ahead or to the left, the door sensor will be set to `0'.  The animat's goal is to move as far across the maze as possible (see Methods). Note that this representation does not show when the animat is stationary (waits) or retraces its path.}
\label{fig:maze_intro}
\end{figure}

The $2^n\times 2^m$ probabilities of an $n$-input and $m$-output state transition table, as well as how each HMG is connected to other gates, is encoded within a genome that, when read by an interpreter, creates the network (see Methods, Text S2 and Figure S1, as well as Ref.~\cite{HintzeAdami2008} for a similar structure). Populations of genomes are evolved using a standard Genetic Algorithm (but without crossover, see Methods). To calculate the fitness of each genome, the controller generated from the sequence is transplanted into the animat shown in Fig.~\ref{fig:robot} and tested on its ability to traverse a maze that consists of repeated vertical walls at varying distance to each other, with a single door placed at random locations within the wall. Within each door, a ``beacon" indicates the direction to follow for the shortest path to the next door, but this information is erased the moment the animat emerges from the door. Thus, in order to use this information, it has to be stored in memory for later usage. The actual maze has at least 26 walls to traverse before the maze repeats. A section of a typical maze along with an adapted animat's  trajectory as well as the states of the memory and motor bits are shown in  Fig.~\ref{fig:maze_intro}. Videos S1 to S3 show several movies that depict the motion of the animat, at different evolutionary stages, traveling through the maze. 

In each of 64 independent evolution experiments, a population of 300 {\em initially random} genomes (encoding random controllers, see Text S2) was evolved for 50,000 generations each. We calculate fitness ($f$) and control fitness $f_{\rm ctrl}$ both for the highest fitness animats at every generation and for genomes on the line of descent (LOD) of the last common ancestor of
the population that existed at generation 50,000 (see Methods). The control fitness  $f_{\rm ctrl}$ tests the performance of the controller on ten randomly generated mazes that the animat has never before encountered (see Methods), in order to test whether the animat evolved the navigation principles or simply adapted to a particular instance of the problem.

\begin{figure}[!ht]
  \centering
  \includegraphics[width=\linewidth]{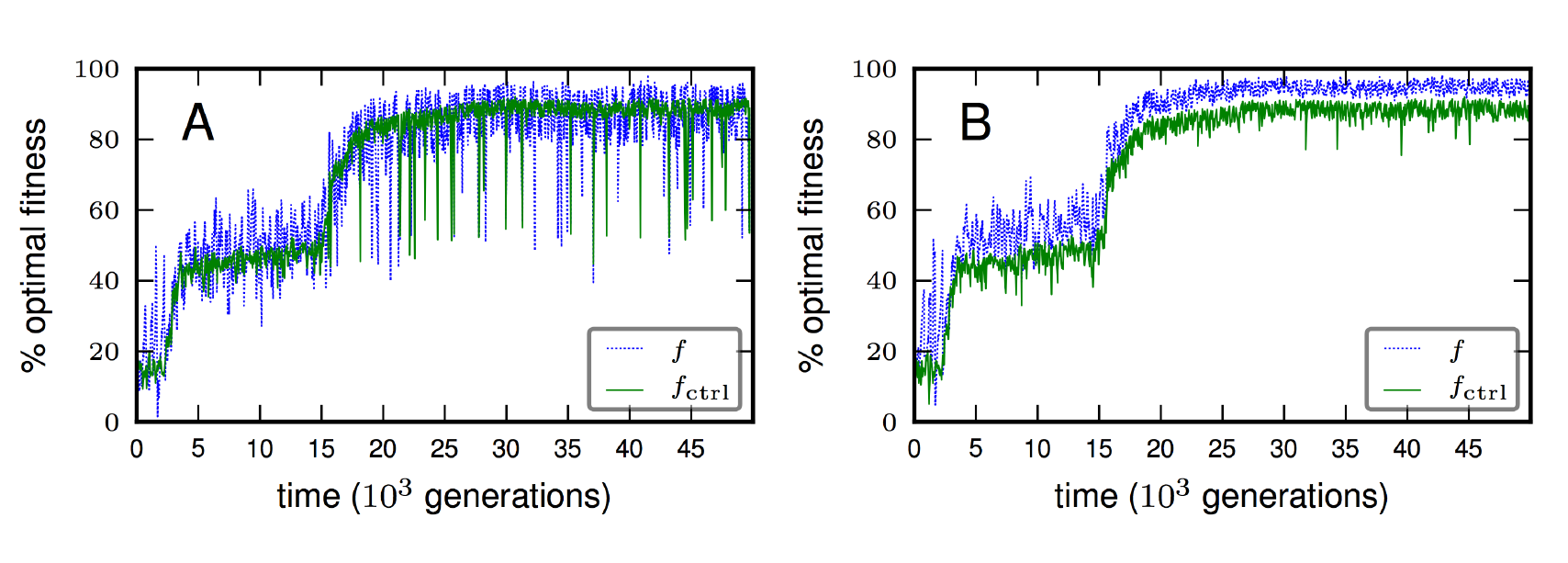}
  \caption{{\bf Fitness evolution on the line of descent and in the population.} {\bf A}: Fitness (blue line) and control fitness (green line) for genotypes on the LOD. {\bf B}: Fitness and control fitness for the same run as shown in (A), but for the fittest individual in the population in each generation. Colors as in (A).}
  \label{fig:Run-52-Fit}
\end{figure}

\begin{figure}[!ht]
\begin{center}
\includegraphics[width=\linewidth]{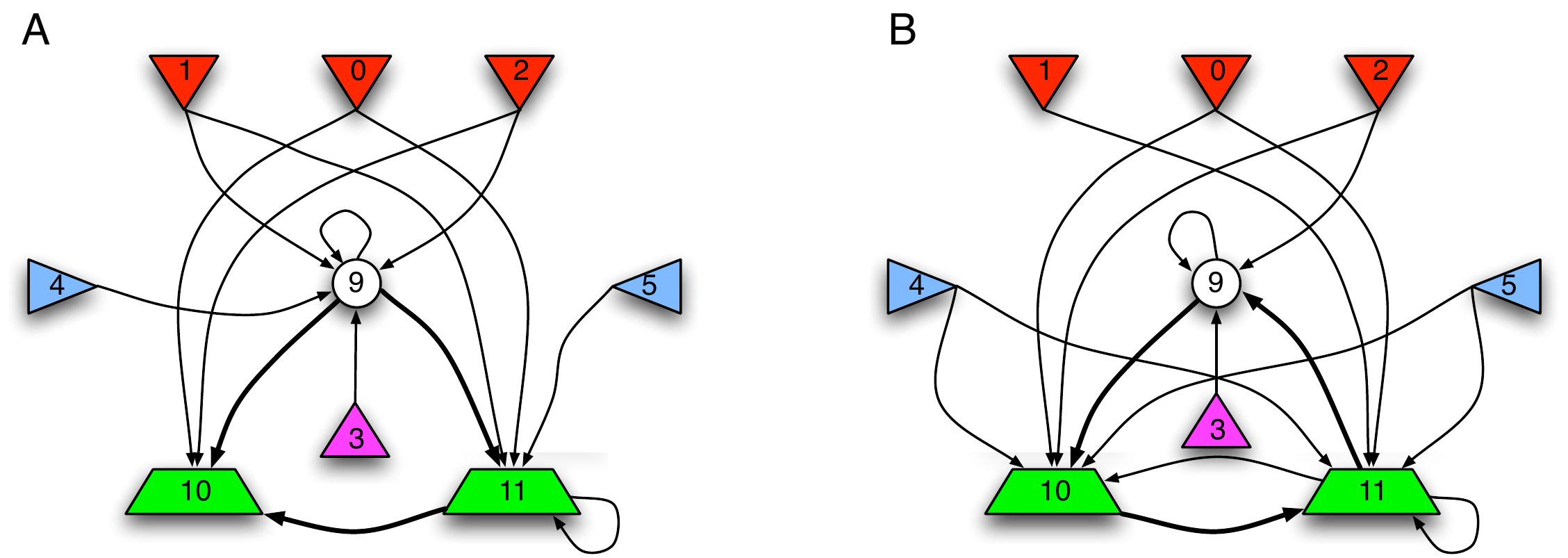} 
\end{center}
\caption{{\bf Two evolved HMG networks.} The shapes represent the 9 Markov variables (bits) at time $t=49,000$ that are active in the network (bits 6, 7, and 8 are connected to the network, but are not functional at generation 49,000 and not rendered here).
The central feed-forward circuit for navigation is rendered in bold arrows.  Color codes and numbering as in Fig.~\ref{fig:robot}. {\bf A}: The network evolved in our focus experiment that achieved 88\% of possible fitness. {\bf B}: Another network that evolved in an independent run, and that implements a variant of the hierarchical temporary memory algorithm that creates an expectation of future sensory signals.  In contrast to the controller that evolved in panel (A), this one uses a feed-back strategy between memory and motors.
This controller achieves 74\% of maximal fitness within a random maze environment.  
}
\label{fig:HMM-Net}
\end{figure}

The LOD  recapitulates the evolutionary history of the population, and allows a reconstruction of the path taken mutation by mutation. Fig.~\ref{fig:Run-52-Fit} shows the evolution of fitness and control fitness for one of 64 experiments [panel (A) shows the fitness on the LOD, while panel (B) shows the corresponding fitness of the best in the population of 300 individuals]. Notice that in
Fig.~\ref{fig:Run-52-Fit}B  the fitness of the fittest individual is almost always larger than the control fitness for the same individual, while the fitness on the LOD instead scatters around the control fitness, as seen in Fig.~\ref{fig:Run-52-Fit}A. There is a good reason for this difference: in any population, an animat can be fit by chance through having correctly ``guessed" the next door position repeatedly. The control fitness removes this element of chance by testing the individual on ten randomly generated mazes. The individuals on the LOD on the other hand are there for a reason and not by chance: their genes have proven themselves in later generations.
In the run depicted in Fig.~\ref{fig:Run-52-Fit} (see also the movies Video S1-S3), the animat evolved a sophisticated (but not perfect) algorithm to navigate the maze, including the use of memory around generation 15,000 to store the doorway bit until the animat reaches the next wall. The wiring diagram of the animat at generation 49,000 is depicted in Fig.~\ref{fig:HMM-Net}A. The animat uses only internal node 9 as memory, whose permanency is ensured using auto-feedback. The other nodes are connected but have no fitness impact whatsoever at this time, as determined by a knock-out analysis (see Methods, data not shown), but may have been useful earlier on. The controller contains a total of 17 HMGs, but only nine HMGs (including two pairs of redundant HMGs) are responsible for this wiring. Of the nine useful HMGs, five have three inputs and one output, the other four HMGs are NOT gates. Note that if more than one HMG output serves as input for another HMG, their values are combined using an OR gate.  
The animat uses the information from the 3-bit retina, the lateral sensors, as well as the conditional information from the door beacon (sensor 3 in Fig.~\ref{fig:HMM-Net}A) effectively by integrating this information within the decision machinery for navigation. The central hub is the network's memory: internal bit 9 is set to 0 if the door beacon is detected in the ``on" state (b3=1) in a doorway, and to 1 if not.  The value of bit 9 is maintained until the animat reaches the next wall. (The value of the door bit itself is erased from the sensor after the animat passes through the door, and therefore cannot be accessed by simply re-reading that value.)  At that point the value of bit 9 determines if the animat goes left (b9=1) or right (b9=0). Once the animat is moving along a wall, bit 9 is set to 1 and the animat continues moving in the same direction keeping in mind the value of the left motor (b11).  In a sense, the motor bit b11 is also used as memory here, as indicated by the auto-feedback. If bit 5 indicates an obstacle to the right, bit 11 is set, which forces bit 10 off in turn.  If bit 4 indicates an obstacle to the left on the other hand, bit 9 is set to 0 which causes bit 11 to turn off and bit 10 to turn on. Once the animat is in front of the next doorway, it moves forward through the door. Thus, we see that this animat effectively uses the integration of different streams of information (door sensor, retina, lateral sensors, and current state of motion) to compute behavior that is appropriate in the given environment most of the time. Reaching 88 \% of ``maximal" fitness is fairly remarkable, as a hand-written optimal controller reaches only 93\% of maximal fitness (data not shown) because we force our controllers to be minimally stochastic. 

In another run that achieved a fitness of 74\%, a related but fundamentally different algorithm evolved to achieve almost the same functionality (wiring depicted in Fig.~\ref{fig:HMM-Net}B). The central part of this algorithm, which is a version of the ``hierarchical temporal memory algorithm"~\cite{HawkinsBlakeslee2004}, is implemented by a feedback loop between the motors 10 and 11 and internal bit 9 (bold arrows in Fig.~\ref{fig:HMM-Net}B), as opposed to the feed-forward loop seen in Fig.~\ref{fig:HMM-Net}A. Because the animat can read from its motor bits, it can keep track of how it is currently moving, and make decisions based on this state as well as the state of the internal variable bit 9. Temporal memory is achieved by creating a basic expectation (bit 9 set to one) of encountering a door beacon that will be pointing it to the left (bit 3=0). If instead it encounters a door pointing to the right (bit 3=1), it changes that expectation (bit 9=0) and maintains it in memory until it moves in the correct (right) direction. Once this happens,  the expectation is changed back to anticipating a beacon pointing it to the left, but the animat does not immediately react to this expectation because bit 9 is ignored as long as the animat moves to the right.

\begin{figure}[!ht]
  \centering
   \includegraphics[width=\linewidth]{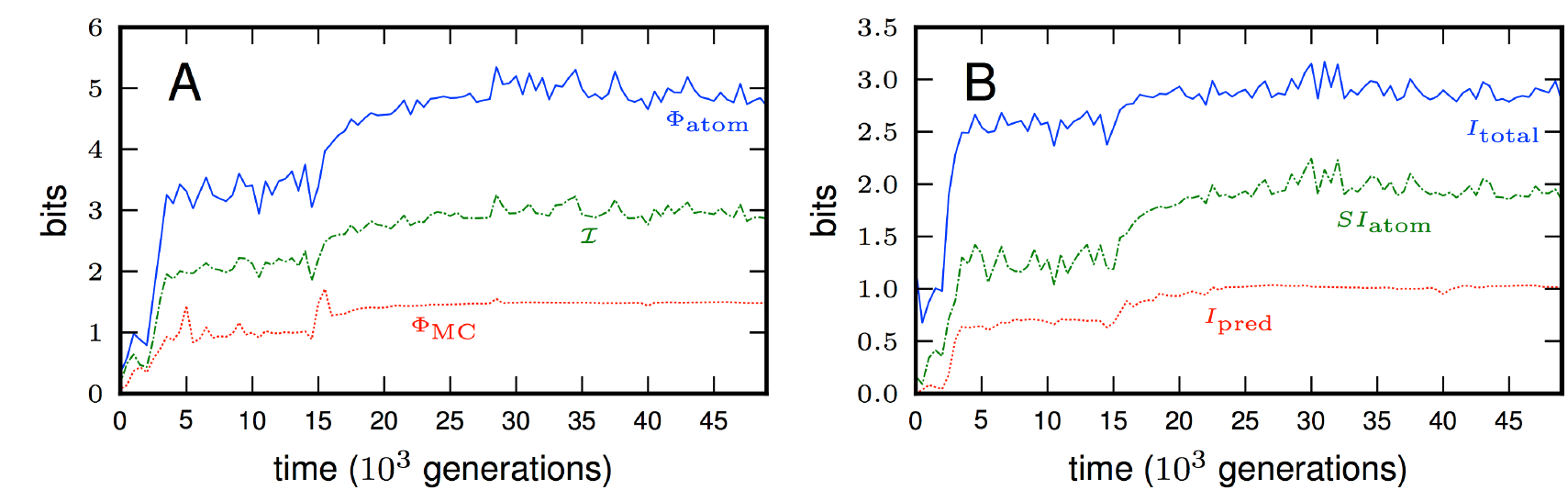}
  \caption{{\bf Information-based measures of complexity.} {\bf A}: Three $\Phi$ related measures of information integration for genomes on the LOD of the same run as shown in Fig.~\ref{fig:Run-52-Fit}. 
Blue line: $\Phi_{\rm atom}$ defined in (\ref{atom}),  green: ${\cal I}$ defined in (\ref{integ}) and red: $\Phi_{\rm MC}$.  
{\bf B}: Three information processing measures for the same experiment as (A): 
Blue: total information $I_{\rm total}$~(\ref{I_tot}),
 green: atomic information $SI_{\rm atom}$~(\ref{iatom}),  and red: predictive information $I_{\rm pred}$~(\ref{ipred}).
}
  \label{fig:Run-52-Infos}
\end{figure}

Let us now look at our information-theoretic constructions as a function of evolutionary time. The quantity $\Phi_{\rm MC}$ is expensive to calculate so they and other measures were calculated along the LOD of each population every 500 generations up to generation 50,000. Each genome was evaluated by testing the controller it spawns for 1,000 world-time steps in 10 control mazes (each tested 10 times, see Methods) in order to even out chance achievements (animates can achieve high fitness by chance due to the stochastic nature of their controllers). We show the evolution of three information integration and three information processing measures over time (for the same run whose fitness evolution is depicted in Fig.~\ref{fig:Run-52-Fit}) in Fig.~\ref{fig:Run-52-Infos}.

As fitness increases, all measures we plot here increase at first, but quickly become stagnant when fitness flattens out (see Fig.~\ref{fig:Run-52-Fit}). Important changes are apparent in all measures when the capacity to use the door beacon for navigation emerges around generation 15,000. To see differences in the measure's abilities to predict fitness, we need to analyze how well these complexity proxies correlate with fitness across our set of 64 runs.

\subsection*{Statistics} 
\begin{figure}[!ht]
\begin{center}
\includegraphics[width=\linewidth]{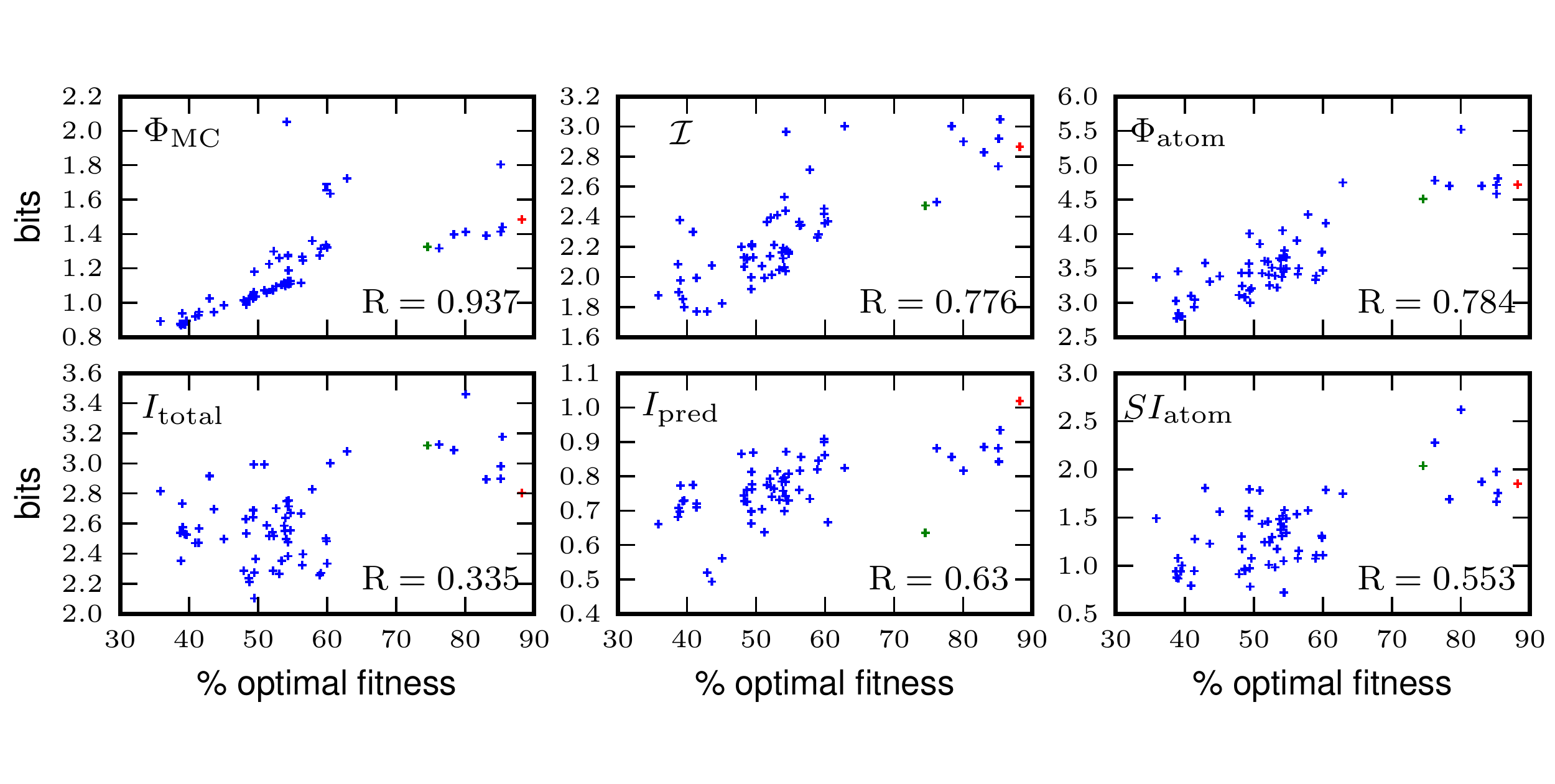}
\end{center}
\caption{{\bf Correlation of information-based measures of complexity with fitness.} $\Phi_{\rm MC}$, ${\cal I}$, $\Phi_{\rm atom }$, $I_{\rm
    total}$, $I_{\rm pred}$, and $SI_{\rm atom}$ plotted against
  $f_{\rm ctrl}$ (as a percentage of optimal fitness) using the final fitness (generation 49,000) on the LOD trajectory for 64 independent runs. $R$ indicates Spearman's rank correlation coefficient. }
\label{fig:MultPhi}
\end{figure}

In order to test whether fitness correlates with a complexity proxy, we calculate the (nonparametric) Spearman rank correlation coefficient of the ``final" fitness (the fitness of the genome at generation 49,000 on the LOD, see Methods) with the value of that variable measured at generation 49,000. We chose generation 49,000 as final time because the organism from this generation is guaranteed to represent the common line of descent of the 300 individuals in any particular run (see Methods).  While we have correlation data of fitness with each variable along the LOD every 500 generations for each run, these points are not independent, and therefore cannot be used in order to assess the statistical significance of the correlation. The correlation of final fitness (across 64 independent samples) with each of the different information-theoretical candidates for functional complexity is shown in Fig.~\ref{fig:MultPhi}. Note that the highest control fitness achieved across the 64 runs is $f_{\rm ctrl}=88.27\pm 0.78$, or almost 90\% of perfect performance (see Methods for our definition of fitness). 
That data point (for the run shown in Fig.~\ref{fig:Run-52-Fit}, giving rise to the controller depicted in Fig.~\ref{fig:HMM-Net}A) is indicated in red in Fig.~\ref{fig:MultPhi}. The run that evolved the controller shown in Fig.~\ref{fig:HMM-Net}B is colored green in Fig.~\ref{fig:MultPhi}. 
\begin{table}[!h]
\caption{Spearman's rank correlation coefficients ($R$) and significance ($p$-value) between different candidate measures of functional complexity with ``final fitness", using the  values achieved at generation 49K of the LOD (an approximation of the most recent common ancestor) for 64 independent runs.}
\label{tbl:corTable}
\begin{center}

\begin{tabular}{ |c |  c c c c c c |}
\hline
 & $\Phi_{\rm MC}$ & $\Phi_{\rm atom }$ & $\mathcal{I}$ & $SI_{\rm atom}$ & $I_{\rm total}$ & $I_{\rm pred}$ \\ \hline
$R$ & 0.937 & 0.784 & 0.776 & 0.553 & 0.335 & 0.63 \\ \hline
$p$ & $4.1 \times 10^{-30}$ & $1.8 \times 10^{-14}$ & $4.8 \times 10^{-14}$ & $2.1 \times 10^{-6}$ & $6.8\times 10^{-3}$ & $2.4\times10^{-8}$ \\ \hline
\end{tabular}
\end{center}
\end{table}

For all measures, we observe positive and highly significant correlations with fitness (Fig.~\ref{fig:MultPhi} and Table 1). The best correlation is achieved for the integrated information measure $\Phi_{\rm MC}$ ($R=0.937$), followed by the information integration across the atomic partition $\Phi_{\rm atom}$ (Spearman's $R=0.784$), while the correlation with $I_{\rm pred}$ is weaker ($R=0.63$). Likewise, $I_{\rm total}$, which does not attempt to separate out the integration of different streams of information correlates only weakly with fitness  ($R=0.335$). The atomic processed information $SI_{\rm atom}$ [Eq.~(\ref{iatom}), $R=0.553$] and the integration ${\cal I}$ both contribute to the strong correlation of $\Phi_{\rm atom}$ with fitness,  as $\Phi_{\rm atom}$  is a sum of ${\cal I}$ and $SI_{\rm atom}$ as per Eq.~(\ref{isum}). The integration measures $\Phi_{\rm atom}$, $\Phi_{\rm MC}$, and ${\cal I}$ also correlate well with each other (data not shown). The difference in the correlation coefficients for  $\Phi_{\rm MC}$ and $I_{\rm pred}$  is highly significant ($p=0$ in a Fisher r-to-z transformation test). 

A clear separation between runs that achieved high ($> 70\%$ of maximal fitness) and low fitness ($< \approx 60\%$) is apparent in Fig.~\ref{fig:MultPhi}, indicating the difference between controllers that can or cannot access the information in the door beacon, which in turn requires the evolution of at least a single bit of memory. However, while it is not possible to achieve fitness in excess of 70\% without using the information from the door beacon, one run utilized this information without exceeding 60\% fitness, as determined via a knock-out analysis of the Markov variables. 

\section*{Discussion}
We have characterized several different information-theoretic measures in terms of their ability to 
reflect the complexity of information processing and integration in discrete dynamical systems. 
In order to discuss non-trivial examples of networks that are functional, we evolved computational networks
that control an animat's behavior in a maze, and tested whether an increase in appropriate behavior is correlated with the putative proxies for complexity. The ``brains'' we evolved capture the essence of what it means to be successful in the maze environment: they can navigate arbitrary mazes of the type they are confronted with, and perform equally well with random versions of mazes that they never encountered during their evolution. In particular, they integrate the sensory information from several sources appropriately, and when they evolve memory they are able to implement it in a variety of ways, including a variant of the hierarchical temporal memory paradigm~\cite{HawkinsBlakeslee2004}. 
We find that a standard measure that has been used to characterize complex robot behavior in the past~\cite{Ayetal2008}, the predictive information $I_{\rm pred}$, usually correlates well with fitness but sometimes fails to do so. We found examples where the failure to be predictive of fitness is associated with the evolution of memory (for example, the run indicated in green in Fig.~\ref{fig:MultPhi}), but also examples where this is not the case (e.g., the run that achieved the highest fitness, shown in red in Fig.~\ref{fig:MultPhi}). 
We hypothesize that when memory emerges, the integration of information from memory with the other signal streams is best reflected by measures of information integration such as $\Phi_{\rm atom}, {\cal I}$, and $\Phi_{\rm MC}$. Indeed, it is possible to show under fairly general assumptions that measures like $I_{\rm pred}$ can maximize fitness under the condition that no other information is used by an agent (such as acquired or inherited information~\cite{RivoireLeibler2011}, see also Text S1). Thus, while we expect that $I_{\rm pred}$ performs worse and worse as a predictor of complex function as more and more memory is utilized for navigation, in some cases $I_{\rm pred}$ turns out to perform very well counter to this expectation. It is currently unclear what is at the origin of this difference in predictive performance of $I_{\rm pred}$. 

That $I_{\rm pred}$ ultimately has to fail as a predictor of fitness when memory is used can be seen in the limiting case of navigating entirely by memory. In that case, any correlation between sensory inputs and motor actions would be purely accidental, in particular if the sensory data that ultimately predict appropriate motor actions are not in the immediate past. However, a non-Markovian version of $I_{\rm pred}$ that takes sensorial data from more distant time steps into account could conceivably perform well even in this case.

On the other hand, measures of information integration could still be elevated even when navigating by memory, as the motor units are driven by streams of information emanating from within, rather than without. However, as sensorial information is not integrated, measures of information integration should be lower when navigating entirely by memory as opposed to navigating via sensors complemented by memory. At the same time, a brain that dreams rather than acts has vanishing predictive information (as the sensor inputs as well as the motor units have vanishing entropy). Yet, integrated information could still be high, and thus reflect complex information processing in the brain even in the absence of behavior. In this respect, measures of integrated information are a good candidate for a quantitative measure of consciousness, as advocated earlier~\cite{Tononietal1994,balduzzi-tononi-08,tononi-08,Tononi2010}. We note, however, that evolving functional networks with high $\Phi$ is not easy. For our 12-bit controllers, $\Phi_{\rm MC} < 2$~bits almost always, and the main complex is significantly smaller than the network size, usually only comprising sensor and motor variables, and occasionally the memory bit when it is used.

Clearly, how useful $\Phi$ is as a measure of brain complexity let alone consciousness rests on testing it on more complex networks that enable complex behavior in simulated environments that are both deep and broad. Evolving networks that rely heavily on memory, and that have the capacity to observe {\em their own state}~\cite{Adami2006} and integrate that information with the sensorial stream, would be particularly useful in this respect. Ultimately, we expect that measures of information integration can then turn into predictors of fitness or function rather than the other way around. Indeed, the functional complexity of biological organisms (measured in terms of fitness) can only be estimated in the rarest of cases when we have a full understanding of what makes an organism successful in its particular niche. 

In future work, we hope to evolve animats in more complex environments that require more broad and versatile use of memory, to thoroughly test the hypothesis that information integration measures outperform pure processing measures such as predictive information in complex tasks. Furthermore, we plan to test whether animats evolve {\em information matching}~\cite{Tononietal1996}, that is, whether the integrated informational structure generated by an adapted complex fits, or ÔmatchesÕ, the informational structure of its environment. As it is possible to determine in detail how information about the world is represented within the Markov brains of these animats, the evolution of such creatures should demonstrate that evolution can move beyond representation-free AI~\cite{Brooks1991} towards autonomous intelligence. 

\section*{Methods}

\subsection*{Agent embodiment}
Of the six sensors shown in Fig.~\ref{fig:robot}, three are obstacle detection bits (binary sensors that
indicate that an obstacle is in front of it (bits 0-2 encode front, left-front, and right-front, respectively), as well as a ``door beacon" (bit 3) that indicates whether the {\em next} opening will be to the right (bit 3=`1') or else in front or left (bit 3=`0') of the opening that the animat is currently passing through. This bit can be used to navigate more successfully in the maze, by keeping this bit in memory and integrating this information with the other sensors. The next opening-direction information is not available after the animat passes through the previous opening. Because the animat cannot turn, it is important to detect whether an animat has hit a lateral wall. Detectors 4 and 5 each return `1' if there is a wall to the left or right respectively.

For example, in Fig.~\ref{fig:maze_intro}, the opening-direction bit (bit 3=`1') in the door just after the starting location indicates that the subsequent opening is to the right. After reaching this door and stepping through it, the sensor bit is set to bit 3=`0' indicating that the next opening is to the left or in front (in this case, in front).  Therefore, efficiently navigating the maze requires memorizing this bit when the animat is in an opening and acting on that information until the animat can see the next opening. 
Two output bits (motors) control each animat's movement: the animat moves right if only bit 10 is on, left if only bit 11 is on, and forward if both are on. The animat has four internal bits (circles 6-9 in Fig.~\ref{fig:robot}) that it can use for information memorization and integration.

\subsection*{Hidden Markov gates}The table depicted within each HMG in Fig.~\ref{fig:hmg} represents the gate's function in terms of a stochastic finite state machine. The binary state of each HMG's inputs corresponds to a row in its probability table.  These probabilities are encoded within the genes that specify the network, as described in Text S2. To determine how those probabilities generate an output from an input, first a random number between 0 and the sum of the elements of that row is generated. Comparing this random number to the cumulative sum of the numbers in that row selects an element in the row whose column index corresponds to the binary state of that gate's outputs. The OR operator is used to combine the outputs from multiple gates which output to the same bit.

\subsection*{Genetic encoding of network structure}
Networks are encoded within circular genomes that are given by a sequence of unsigned characters [0,255]. Each gene encodes a single HMG  and its  connection to other gates via the Markov variables, as well as the state-transition probabilities that define the gate. Details about the interpretation of the genome and its translation into a network are given in Text S2. Each HMG can have at most 4 inputs, and at most 3 outputs. If more than one HMG writes into a single Markov variable, these outputs are combined via an OR operation but we allow at most 3 write-attempts into a single Markov variable. If in the sequential interpretation of the genome an HMG requests to write to a variable that already has three connections, that HMG's connection will instead be routed to the nearest available variable. The same restrictions exist if an HMG tries to read from a variable that already has 3 read connections.

\subsection*{Fitness calculation}
The animat's fitness in a maze $m$ is determined by:
\begin{equation}
g(m) = \sum_{t=0}^{T} \left(\frac{D-d_t}{D} + L_t\right)
\end{equation}
where $T$ is the number of time steps ($T=300$),  $d_t$ is the shortest path distance to the last doorway in the maze from the animat's position at time $t$, $D$ is the maximum shortest path from all locations in the maze to the last doorway, and $L_t$ is the number of times the animat has passed the last doorway. The maze environment is periodic so that if the animat goes past the end of the maze, the environment is the same as the beginning of the maze.

The stochastic nature of the controller implies that the $g(m)$ measured in one run through a single maze $m$ may not be a reliable estimate of the genome's fitness because chance decisions could lead to either too high or too low fitness. We therefore define the animat's selection fitness by:
\begin{equation}
f(m) = \left(\prod_{i=1}^{10}\frac{g^{(i)}(m)}{g_{\rm opt}(m)}\right)^{\frac{1}{10}} \label{eq:fitness}
\end{equation}
where $g^{(i)}(m)$ is the $i^{\rm th}$ stochastic realization of the animat's fitness in maze $m$, and $g_{\rm opt}(m)$ is the maximum fitness attainable in that maze. The geometric mean of 10 evaluations helps to ensure the reproducibility of the animat's fitness, and make it a better predictor of the long-term success of the lineage it represents. 

In order to ensure that the genomes have evolved the ability to navigate through {\em general} mazes of this type (rather than adapting to a single particular maze), a set of 10 control mazes are used to calculate the {\em control fitness}:
\begin{equation}
f_{\rm ctrl} = \frac{1}{100}\sum_{m=1}^{10}\sum_{i=1}^{10} \frac{g^{(i)}(m)}{g_{\rm opt}(m)} \label{eq:ctrlFit}
\end{equation}
The control fitness uses the arithmetic rather than the geometric mean in order to better track performance. The geometric mean in Eq.~(\ref{eq:fitness}) allows for the elimination of controllers that ever completely fail at a single instance (as fitness is then multiplied by zero). The arithmetic mean in Eq.~(\ref{eq:ctrlFit}) is a better numerical indicator for the power of the strategy, as a single failure does not result in a vanishing control fitness. 

\subsection*{Evolution and Genetic Algorithm} 64 populations of 300 individuals were evolved for 50,000 generations. For the purpose of selection, a single maze was randomly generated for each run, given a set of boundary conditions.  Every 100 generations a new maze was generated for each run so that the animats would not adapt to a specific instance of the problem. To implement selection,  the top three individuals from each generation (the elite) were copied into the next generation without mutation, unless their fitness was determined to be zero after re-testing. The remaining places in the population were filled by roulette-wheel selection with mutations~\cite{Michalewicz1999}. This implies that the number of offspring that any parent places into the next generation is proportional to the relative fitness advantage (or disadvantage) it holds with respect to the average population fitness. However, no individual could place more than 10 offspring into the next generation. The genomes (described in Text S2 and Fig. S1) were changed via a variety of processes from generation to generation. Single loci were copied with a probability of $\mu_{\rm site copy}=0.025$, deleted with probability $\mu_{\rm site del}=0.05$, a random value inserted after a loci with probability $\mu_{\rm site ins}=0.025$, replaced with a uniformly drawn random number $\in[0,255]$ with probability of $\mu_{\rm site uniform}=0.05$, or increased/decreased by a random number $\in[-10,10]$ (restricted to the range $[0,255]$ if necessary) with probability of $\mu_{\rm site\ up/down}=0.05$. Whole genes where duplicated with $\mu_{\rm Gdup}=0.005$, deleted with $\mu_{\rm Gdel}=0.01$, and a random gene inserted with $\mu_{\rm Gins}=0.005$.

Finally, all mutation rates were normalized such that the whole genome mutation rate is equal to one change per genome per generation on average. This has the consequence that the ``expressed" genome fraction (fraction with functioning start codon giving rise to HMGs connected to the main network) decreases with evolutionary time. Around 50,000 generations, the amount of expressed genes is of the order of 15\% of the total genome size (on average about 200 of 3,000 loci are expressed in an evolved genotype).

\subsection*{Knock-out analysis}
In order to determine the importance and role of individual variables in the brain's operation, we perform ``knock-outs" on the variables to test their effect on the Markov animat's performance. Four types of per-bit knockouts were used: replace the value that the variable takes on by `always read 0', `always read 1', `always write 0', and `always write 1'. Some brains use variables with fixed values on purpose, in order to select certain rows from the probability tables with certainty. Such variables can be detected when only one of the two read-knockouts (read-zero or read-one) reduce the fitness of the controller. Variables that actually store and/or process information will lead to reduced fitness by both knockouts. Motor variables that are not read from are unaffected by the read knockouts but are affected by the write knockouts. Similarly, write-knockouts from sensor variables do not affect fitness, while read-knockouts do.

To determine the function of individual HMGs, first each HMG was deleted from the controller to see if it changed the fitness. This identified unique important HMGs, but sometimes the results were masked by redundant HMGs. Then, each entry in the probability table for each HMG was ``knocked out" by replacing the corresponding allele by zero or 255 [see Eq.~(1) of Text S2 for the effect of this replacement]. This data combined with the input distribution for each HMG was used to determine the role of any particular HMG in the brain, and how it worked together with the other HMGs to control the animat.

\subsection*{Line of descent}

For each run the line of descent (LOD) was obtained~\cite{Lenskietal2003} by tracing back the fittest organism in the population backwards towards the randomly constructed ancestral sequence used to begin each experiment (encoding on average 12 HMGs, see Text S2). Seen from the point of view of the ancestral sequence, each following generation creates a branching tree with some lines eventually becoming extinct and other branches surviving. Because we simulate an asexual population in a single niche, only a single line of descent can ultimately remain because of competitive exclusion between members of the same species~\cite{Hardin1960}. This line can be identified from following the lineage back from {\em any} of the 300 organism present in the final generation (generation 50,000) back to the origins (300 individual lines of descent). Going back ten generations, say, to 49,990, there will be fewer lines because some lines coalesced going backwards (branched going forward). The further backwards one moves on this ``tree of descent", the more lines coalesce until the last common ancestor (LCA) of the entire population that was alive at generation 50,000 has been reached. Because of the single-niche environment, the 300 lines coalesce very quickly, and are virtually guaranteed to have coalesced to a single line when going back to generation 49,000, which is the ``final" generation we study in our simulations, and defines the ``final fitness". The organism at generation 49,000 of the LOD is not guaranteed to be the LCA, but it is guaranteed to be the ancestor of all organisms present in the final generation. Thus, the LOD records the evolutionary history of the experiment mutation by mutation, and allows us to reconstruct the evolutionary path that led to the adapted type. Fitness as well as complexity measures were calculated for organisms on the LOD every 500 generations.

\section*{Acknowledgments}
We would like to thank David Balduzzi, Virgil Griffith, Nikhil J. Joshi, Sang Wan Lee, and Jory Schossau, and acknowledge detailed and insightful reviewer comments. 
This work was funded in part by the Paul G. Allen Family Foundation, by the National Science Foundation's Frontiers in Integrative Biological Research grant FIBR-0527023,  NSF's BEACON Center for the Study of Evolution in Action under contract No. DBI-0939454, and by the WCU (World Class University) program through the National Research Foundation of Korea funded by the Ministry of Education, Science and Technology (R31-10008).

\bibliography{Phi}

\newpage


\clearpage

\clearpage
\renewcommand{\theequation}{S\arabic{equation}}
\setcounter{equation}{0}
\newpage
\setcounter{page}{1}
\pagestyle{empty}
\section*{Supporting Information} 

\setcounter{figure}{0}
\renewcommand{\thefigure}{S\arabic{figure}}
\begin{figure}[!ht] 
   \centering
   \includegraphics[width=0.9\linewidth]{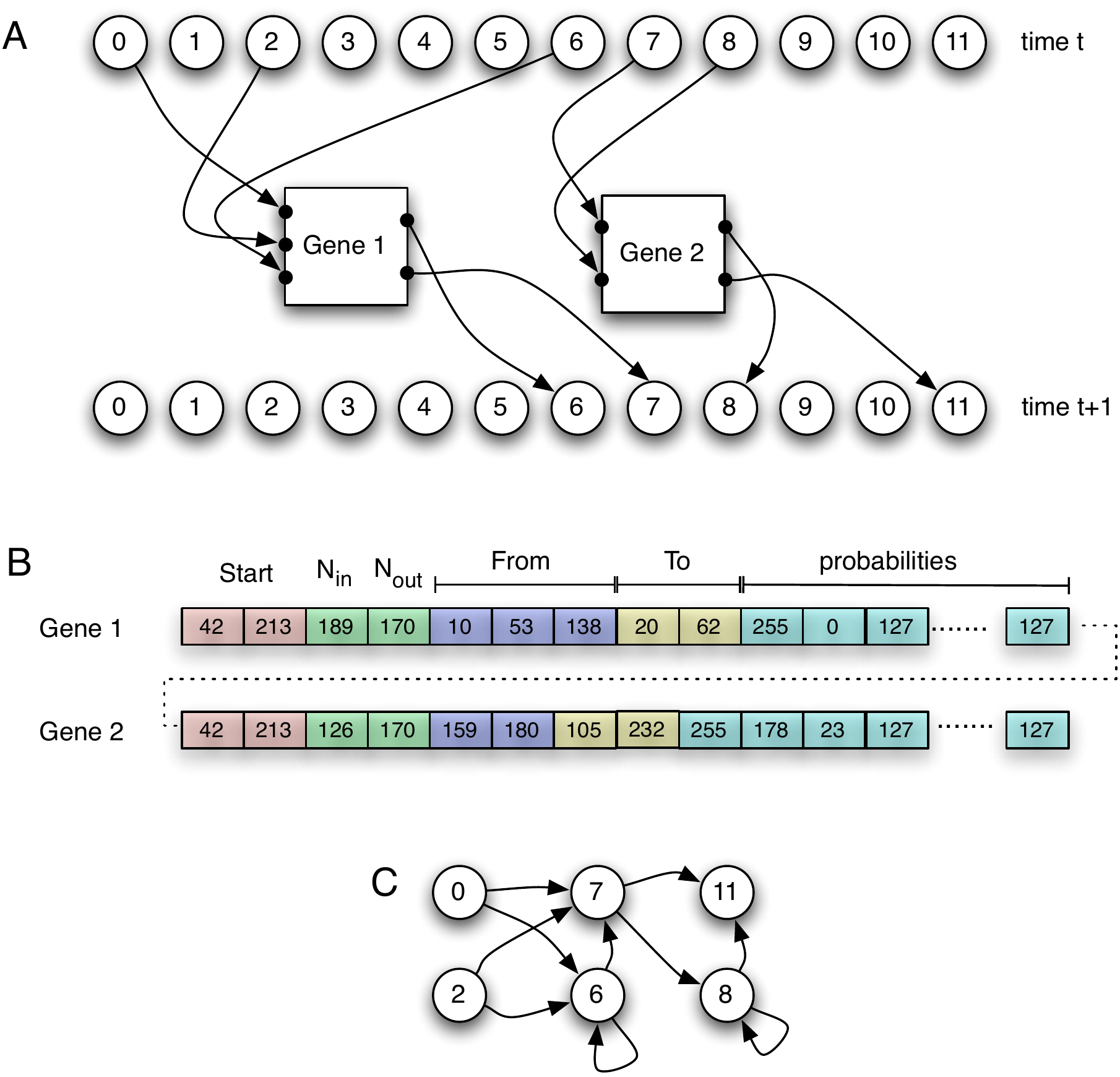} 
   \caption{{\bf Genetic encoding of animat controllers.} {\bf A}: In this example, two HMGs encoded by two genes can read from and write to several of the 12 Markov variables, indexed 0-11. The top row shows the Markov variables at time $t$ that the HMGs can read from while the row below shows how the HMGs write into those variables to update their state at $t+1$. 
{\bf B}: The genome is a circular sequence of loci that carry unsigned integers ${\tt allele}\in[0,255]$  and encode the input output structure of each HMG as well as the connectivity between them and the state transition tables that determine each HMG's function. Colors denote different functional sections of the gene. {\bf C}:  Causal influence of the Markov variables induced by the two HMGs. Presence of an arrow between variables $a$ and $b$ implies that $a$ may change the state of $b$ in a single time step. Absence of an arrow implies that the variables cannot influence each other within a single time step.} 
   \label{fig:genome}
\end{figure}


\subsection*{Video S1}
This movie shows the trajectory of an evolved animat traveling through the maze after 2,000 generations of evolution in the top panel, and
the inner workings of its Markov network brain in the lower panel. At this point in evolutionary history, the animat has learned to move forward whenever it stands in front of an opening, but otherwise performs a random walk. The fitness at this time point is $15.8 \pm 0.6\%$ of maximal.

The animat in the maze is depicted with a triangle, and the trail it leaves reflects the activation pattern of its four internal nodes and its
motor outputs, as described in Fig.~\ref{fig:maze_intro}.  The brain state (lower panel) shows all HMGs (U0-U10) and the probabilities in the
state-transition tables as percentages (colored in shades of gray). Input bits (labeled iB) and output bits (labeled oB) are green
if true and blue if false.  The red element in each table indicates the element of the table selected at that time step based on the values of
the input bits and the probabilities in that row. In other words, a table element turning red indicates which state of the HMG was selected as a function of the input. This is akin to a pattern of neuronal firings as a function of the inputs.

\subsection*{Video S2}
The trajectory and brain states of an evolved animat at generation 14,000. At this point, the animat has acquired the capacity to maintain a direction of travel and move opposite to the direction indicated by the lateral contact sensor. Its movement with respect to the door opening is still random. The fitness at this time point is $47.9\pm0.7\%$ of maximal.
\subsection*{Video S3}
The trajectory and brain states of an evolved animat at generation 49,000. By this time, the animat has evolved the capacity to use the information provided by the door beacon by storing it in bit 9, and move purposefully in the indicated direction after emerging from the previous door.  Because of its high fitness, the animat traverses the maze five times, but does not always take the same trajectory every time, illustrating the stochasticity of its decisions. The fitness at this time point is $88.2\pm 0.7\%$ of maximal.

\pagebreak

\subsection*{Text S1. Relationship between different forms of  $I_{\rm pred}$}
Our definition of $I_{\rm pred}$ as the shared entropy between sensor variables at time $t$ and actuator variables at time $t+1$ ostensibly differs from that of Bialek et al.~\cite{Bialeketal2001} and \cite{Ayetal2008},  as well as from measures quantifying the value of information used by Rivoire and Leibler~\cite{RivoireLeibler2011}. This difference is due to the particular way in which our animats interpret sensorial signals via their action variables and the dynamics of how the animat and the environment are updated, but are otherwise directly comparable. Bialek et al.~\cite{Bialeketal2001} define $\ipred$ as the shared entropy between the states of a data stream $X$ in the past and in the future, which for environments with the Markov property reduces to 
\be
\ipred=I(X_t:X_{t+1})\;, \label{ip1}
\ee
which is the shared Shannon entropy between subsequent states of the data stream.
Ay et al.~\cite{Ayetal2008} adapted this measure to the sensor-action loop of an autonomous robot where motor variables $Y_t$ affect the sensed variables $X_{t+1}$ one update later, rewriting the predictive information Eq.~(\ref{ip1}) in terms of sensor and motor variables as
\be
\ipred=I(Y_t:X_{t+1})\;. \label{ip2}
\ee
This differs from our Eq.~(\ref{ipred}) when identifying $X_t\equiv S_t$ and $Y_t\equiv R_t$ because of different ways in which the dynamics
of the systems are updated. In  \cite{Ayetal2008}, the authors advance the time counter when the actions of the controller are applied to the environment, so that
\be
x_{t+1}=F(x_t,y_t)+\xi_{t+1}\;, \label{update}
\ee
where $\xi_{t+1}$ is a Gaussian white noise term and $F(x,y)$ is a function mapping old sensor and motor variables to their updated values. Instead, we advance the time counter when the motor variables are updated based on the sensed environment:
\be
y_{t+1}=G(x_t,y_t)
\ee
with a different update function $G(x_t,y_t)$ that is optimized by evolution and contains stochastic effects from the HMGs. With this updating scheme, the sensed values are the cause and the changed motor variables are the effect while using the update scheme of Ref.~\cite{Ayetal2008}, the motor variables are the cause and the change in sensor values is the effect. Both versions, however, capture the predictive information. Note that when $F(x_t,y_t)=x_t$ in (\ref{update}) as in Ref.~\cite{Ayetal2008}, expression (\ref{ip2}) reduces to (\ref{ip1}). 

Rivoire and Leibler discuss simple models of populations that use environmental states $x_t$ to optimize their growth. In case no information is inherited (or remembered), the fitness of the population is maximized where the Shannon information $I(X_t:Y_t)$ is maximal. Here, the states $y_t$ are environmental signals that the agent perceives {\em and uses} in order to survive optimally in the environment. In that respect, they correspond to those environmental cues that had an effect on the behavior of the agent; thus they are best described by motor variables at time $t+1$ (sensed values that do not affect the motors could as well have been random). Thus, in the absence of inherited information or memory, the Shannon information $I(X_t:Y_t)$ is equivalent to our predictive information Eq.~(\ref{ipred}) in the main text. If memory or other information influences the actions of the agent, $I(X_t:Y_t)$ no longer maximizes the fitness. In a simple model where agents act optimally given the environment, the fitness is maximal at high $I(X_t:Y_t|X_{t-1})$, but if the strategy is non-optimal (as in the case we discuss here), no general expression can be given~\cite{RivoireLeibler2011}.
  
\pagebreak
\subsection*{Text S2. Genetic encoding of network structure and function}
Hidden Markov Gates (HMGs) are encoded by genes that specify where their inputs come from (what Markov variable is read at time $t$) and where the gate writes to, thus updating that Markov variable at time $t+1$. The 12 (binary) variables that we use can be depicted at a point in time $t$ as in Fig.~\ref{fig:genome}A (top). Because HMGs can read or write from these variables, write events from multiple HMGs have to be resolved in order for the variable to take on a unique state at each $t$. This is achieved by combining the inputs to a variable via an OR operation. We allow at most 3 simultaneous write attempts into each variable. An HMG that attempts to write into a variable that already has 3 connections to it has its connection attempt redirected to the nearest variable that has open slots remaining (such redirects are rare). Before the Markov variables are updated, they are all cleared (internal variables as well as sensors and actuators), so that no information can be stored by simply not writing into a variable before it is read. Rather, memory has to emerge using the computational structure of the network. In this sense, the 12 Markov variables are completely passive conduits for information processing. 

An HMG is encoded by a gene that is identified by a ``Start" sequence given by the specific alleles `42' followed by `213' (see Fig.~\ref{fig:genome}B). The start signal is chosen so as not to interfere with common alleles appearing in the genome, such as the numbers `0', or `255'. The exact ``start codon" (42,213) is therefore rare, occurring by chance only once every $2^{16}$ pairs. The next two loci encode the number of inputs $N_{\rm in}$ and $N_{\rm out}$ of the gate, by converting $N=\lfloor\frac{\tt allele}{\lfloor255/N_{\rm max}\rfloor}\rfloor$, where $N_{\rm max}$ is the largest number of inputs or outputs that any HMG can have. In the present implementation, $N_{\rm max}=4$ for inputs and $N_{\rm max}=3$ for outputs. 
The next $N_{\rm in}$ slots encode the identifier of the Markov variable that the HMG should read from, where identifier=$\lfloor \frac{{\tt allele}\times 12}{255} -\frac12\rceil$ and the symbol $\lfloor \cdot\rceil$ indicates the nearest integer. However, if a variable already has 3 or more HMGs reading from it, the connection is rerouted to the nearest available variable. Thus, alleles 10, 53, and 138 in the ``From" block specify that the HMG encoded by gene 1 in Fig.~\ref{fig:genome}B will read from variables 0,2, and 6.
The next $N_{\rm out}$ slots specify the identifier of the Markov variable that the HMG should write to where identifier=$\lfloor \frac{{\tt allele}\times 6}{255} + 5.5\rceil$. This ensures that the smallest identifier that a variable can write to is 6, so that an HMG can never write into a sensor.  Again, not more than 3 HMGs can write into the same variable, so the connection is routed to the nearest available variable instead. According to these rules of translation, alleles 20 and 62 encode the identifiers 6 and 7 that this gate writes into. 

The probabilities that specify the function of the gate are encoded in $2^{N_{\rm in}}\times2^{N_{\rm out}}$ loci following the ``From" and ``To" blocks. These probabilities are determined by the relative value of alleles in each row of the transition table (see Fig.~\ref{fig:hmg} as an example). So, the four alleles (127,127,127,127) in the $i$th row of a table with two inputs encode the probabilities $(p_{i1},p_{i2},p_{i3},p_{i4})=(1/4,1/4,1/4,1/4)$, but the vector (255,255,255,255) encodes the same probabilities. Generally,
\be
p_{ij}=\frac{1+{\tt allele}_{ij}}{\sum_{j=1}^{2^{N_{\rm out}}}(1+{\tt allele}_{ij})}\;.
\ee
We add 1 to each allele in order to ensure that no probability is ever exactly zero, and in particular that the row sum is never zero. If a mutation changes $N_{\rm in}$ or $N_{\rm out}$ or both, the parser will interpret a sufficient number of subsequent loci as probabilities to fill up all the slots in the HMG table (see Fig.~\ref{fig:hmg}).  This may happen by overlapping an adjacent gene, and because genomes are circular the parser will always succeed in assigning values.

Parsing these genes creates a network of HMGs that can be rendered either as a network of HMGs  or else as a network that describes how the Markov variables interact with each other as in Fig.~\ref{fig:genome}C. As an ancestral genome, we use a randomly generated genome of about 12 genes. For these genes, we randomly create the input/output connections, and fill in the probability tables (whose sizes are calculated based on the number of inputs and outputs created) using uniformly distributed random numbers $\in[0,255]$.

\subsection*{Text S3. Relationship between $EI$ and $SI$}
In the main text we defined two measures of information integration across a partition that we called $EI$ [Eq.~(\ref{EI})] and $SI$ [Eq.~(\ref{SI})] which we repeat here:
\be
SI(X_0\to X_t|P)&=&I(X_0: X_t)-\sum_{i=1}^k I(P_0^{(i)}:P_t^{(i)})\;, \label{SI1}\\
EI(X_0\to X_t|P)&=&\sum_{i=1}^kH(P_0^{(i)}|P_t^{(i)})-H(X_0|X_t)\;. \label{EI1}
\ee
In this section, we derive the relationship between these two measures. We begin with arbitrary probability distributions ${\rm Pr}(X_0=x_0)$ and ${\rm Pr}(X_t=x_t)$, and first calculate $I(X_0:X_t)$ defined as 
\be
I(X_0:X_t)=H(X_0)-H(X_0|X_t) \label{q1}
\ee
where
\be
H(X_0)=-\sum_{x_0}p(x_0)\log p(x_0) \label{h0}
\ee
and 
\be
H(X_0|X_t)=-\sum_{x_0,x_t}p(x_0,x_t)\log p(x_0|x_t) \label{h0j}
\ee
where $p(x_0|x_t)=p(x_0,x_t)/p(x_t)$ is the conditional probability to have observed state $x_0$ given that we observed state $x_t$ $t$ time steps later. Of course, Eqs.~(\ref{q1}) and Eq.~(2) of the main text are equivalent on account of (\ref{h0}) and (\ref{h0j}).  
The relationship (\ref{q1}) also holds for each of the $i$ parts of a partition:
\be
I(P_0^{(i)}:P_t^{(i)})=H(P_0^{(i)})-H(P_0^{(i)}|P_t^{(i)})\;. \label{q2}
\ee
If we insert (\ref{q1}) and (\ref{q2}) into (\ref{SI1}) we obtain
\be
SI(X_0\to X_t|P)&=&H(X_0)-H(X_0|X_t)-\sum_{i=1}^k \left[H(P_0^{(i)})-H(P_0^{(i)}|P_t^{(i)})\right]\\
&=&-H(X_0|X_t)+\sum_{i=1}^kH(P_0^{(i)}|P_t^{(i)})+H(X_0)-\sum_{i=1}^kH(P_0^{(i)})\;. \label{si2}
\ee
Together, the first two terms in Eq.~(\ref{si2}) are $EI$ in Eq.~(\ref{EI1}). The last two terms together are the negative of the (positive) integration ${\cal  I_P}(X_0)$ across partitions
\be
 {\cal I_P}(X_0)=\sum_{i=1}^kH(P_{0}^{(i)})-H(X_{0})\;.\label{integ0}
\ee 
This integration across partitions is a generalization of the quantity introduced in Eq.~(\ref{integ}), but at step $t=0$:
Thus:
\be
SI(X_0\to X_t|P)=EI(X_0\to X_t|P)-{\cal I_P}(X_0)\;.
\ee
For a maximum entropy distribution ${\rm Pr}^{\rm max}(X_0)$ (all states appear with equal probability), all possible partitions must also be uniformly distributed as there can be no correlations between them. Thus, for ${\rm Pr}^{\rm max}(X_0)$ (but only for this distribution)
\be
{\rm Pr}^{\rm max}(X_0)=\prod_{i=1}^k {\rm Pr}^{\rm max}(P_0^{(i)})\;.
\ee
This implies that 
\be
H^{\rm max}(X_0)=\sum_{i=1}^k H^{\rm max}(P_0^{(i)})\;.
\ee
In that case, the integration (\ref{integ0}) vanishes, and $EI$ equals $SI$. The same observation was made by Barrett and Seth in equations (25) and (26) of Ref.~\cite{BarrettSeth2011}.

\pagebreak

\end{document}